\documentstyle[epsfig,preprint]{aastex}

\newcommand \lta {\mathrel{\vcenter
     {\hbox{$<$}\nointerlineskip\hbox{$\sim$}}}}
\newcommand \gta {\mathrel{\vcenter
     {\hbox{$>$}\nointerlineskip\hbox{$\sim$}}}}
\newcommand \m {M$_\odot$}
\newcommand \mm {M_\odot}
\newcommand\cms{cm~s$^{-1}$}
\newcommand\grs{$\gamma$-rays}
\newcommand\grb{$\gamma$-ray burst}
\newcommand\grbs{$\gamma$-ray bursts}

\begin{document}
\title{Asymmetric Supernovae from Magneto-Centrifugal Jets}
\author{J. Craig Wheeler $^1$, David L. Meier$^{2}$, James R. Wilson$^3$}
\affil{$^1$Astronomy Department, University of Texas, Austin, Texas 78712;
 wheel@astro.as.utexas.edu}
\affil{$^2$Jet Propulsion Laboratory, 4800 Oak Grove Dr., Pasadena,
CA 91109; dlm@sgra.jpl.nasa.gov}
\affil{$^3$Lawrence Livermore National Laboratory, Livermore, CA
94551-9900; wilson@ricker.llnl.gov}

\begin{abstract}

Strong toroidal magnetic fields generated in stellar collapse can 
generate magneto-centrifugal jets in analogy to those found in 
simulations of black hole accretion.  Magneto-centrifugal jets may 
explain why all core collapse supernovae are found to be substantially 
asymmetric and predominantly bi-polar.  We describe two phases: 
the initial LeBlanc-Wilson jet and a subsequent protopulsar or 
toroidal jet that propagates at about the core escape velocity.  
The prompt LeBlanc-Wilson jets will produce an excess of neutron-rich 
matter and hence cannot be the common origin of supernova explosions;
similar, but less severe problems arise with the protopulsar jet
that may be alleviated by partial evacuation along the axis by 
rotation.  The jets will produce bow shocks that tend to expel matter, 
including iron and silicon, into equatorial tori.  This may help to 
account for observations of the element distribution in Cas A.

A magnetic ``switch'' mechanism may apply in rare instances (low density 
and large magnetic field), with subsequent increase in the speed and 
collimation of the toroidal jet.  The conditions that turn the magnetic 
switch ``on'' would yield a jet that propagates rapidly and with small 
opening angle through the star, depositing relatively little momentum.
The result could be enough infall to form a black hole.   A third, 
highly relativistic jet from the rotating black hole could catch up 
to the protopulsar jet after it has emerged from the star.  The 
interaction of these two jets plausibly could be the origin of the internal
shocks thought to produce \grbs\ and could explain the presence of
iron lines in the afterglow.  Recent estimates that typical \grb\ energy
is $\sim 3\times10^{50}$ erg imply either a very low efficiency for 
conversion of rotation into jets by the Blandford-Znajek mechanism, or 
a rather rapid turnoff of the jet process even though the black hole 
still rotates rapidly.
  
Magnetars and ``hypernovae'' might arise in an intermediate
parameter regime of energetic jets that yield larger magnetic
fields and provide more energy than the routine case, but that are
not so tightly collimated that they yield failed supernova.

\end{abstract}


\newpage

\section{Introduction}
 
The problem of core-collapse has been with us for over 40 years
(Hoyle \& Fowler 1960).
Immediately after the discovery of pulsars, it was reasonable
to explore the issue of whether or not the rotation and
magnetic fields associated with pulsars could be a significant
factor in the explosion mechanism (Ostriker \& Gunn 1971;
Bisnovatyi-Kogan 1971; Bisnovatyi-Kogan \& Ruzmaikin 1976; Kundt
1976).  With typical dipole fields of
$10^{12}$ G and rotation periods of several to several
tens of milliseconds implying electrodynamic power of only
$\sim 10^{44-45}$ erg s$^{-1}$, a strong robust explosion seemed
unlikely.  Several factors have led to a need to re-examine
this conclusion.  The principle one is the accumulating
evidence that core collapse supernovae are distinctly asymmetric.
Aside from its famous rings, HST observations of SN~1987A resolving the
debris show that the ejecta are asymmetric with an axis that
roughly aligns with the small axis of the rings (Pun
et al. 2001, Wang et al. 2002).  Chandra X-ray Observatory 
(CXO) observations of Cas A show that
the jet and counter jet and associated structure are
observable in the X-ray (Hughes et al. 2000; Hwang et al. 2000)
as well as the optical (Fesen \& Gunderson 1996; and references
therein).  The most direct evidence bearing on this
topic is from supernova spectropolarimetry which shows that substantial
asymmetry is ubiquitous in core-collapse supernovae, and that a
significant portion show strong evidence for a single,
wavelength-independent axis of symmetry (Wang et al. 1996; Wang et al.
2001).  Many, even perhaps most, core-collapse supernovae are bi-polar
(Wang et al. 2001, 2002).

The strength of the asymmetry observed with polarimetry
is higher (several \%) in supernovae of
Type Ib and Ic that represent exploding bare non-degenerate
cores (Wang et al. 2001).  The degree of asymmetry also rises
as a function of time for Type II supernovae that have
retained their hydrogen envelopes (from $\lta$ 1\% to $\gta$ 1\%)
as the ejecta expand and one looks more deeply into
the core material (Wang et al. 2001a; Leonard et al. 2000, 2001).
Both of these trends suggest that it is the
inner machine, the core collapse mechanism itself, that is
responsible for the asymmetry.  The observed polarization
requires significant asymmetry; axis ratios exceeding 2 to 1.
To impose the observed strong asymmetry in the final homologous
expansion it is plausible that an axial flow must be established
and maintained for at least several dynamical time scales
of the outer mantle/envelope.
This is the operational definition of a jet.

It also is significant that CXO observations
have determined that jets are routinely associated with pulsars,
not only in binary systems like SS 433, but also isolated objects
like the Crab and Vela pulsars (Weisskopf et al. 2000; Helfand,
Gotthelf, \& Halpern 2001).  One interpretation of these observations
is that the present-day jets in these young objects are vestiges of a
much more powerful MHD jet era that occurred in the first few seconds
of the protopulsar phase when the compact objects were still inside
their progenitor stars as they first attain nuclear densities. During
this period the transient values of the magnetic field and rotation
could have greatly exceeded those observed today, and indeed, as we suggest
here, {\em the reduction of the rotation rate and magnetic field
to their present-day values could have been the
process of energy release that powered the initial explosion}.

Asymmetries associated with neutrino emission may produce dynamical
asymmetries, but it is not clear that they can account for the
polarization observations.  Neutrino
asymmetries may yield a short-lived, essentially impulsive effect
(Shimizu, et al. 1994; Burrows \& Hayes 1996; Fryer \& Heger 2000;
Lai et al. 2001).  Expansion and transverse pressure gradients can
wipe out transient asymmetries before homologous expansion
is achieved (Chevalier \& Soker 1989).  ``Finger'' asymmetries might
be preserved, but it is unclear that they can reproduce the common
feature of a single symmetry axis that is substantially independent
of space and time (Wang et al. 2001, 2002).
Sufficient neutrino impulse might be delivered
to the neutron star to yield a substantial runaway velocity
(Burrows \& Hayes 1996; Spruit \& Phinney 1998), but it is difficult
to see how this  impulse can be communicated in a substantial and
permanent way to  the final ejecta trajectories.

Highly-resolved, fully three dimensional,
adaptive grid numerical calculations (Khokhlov 1998) have,
however, established
that non-relativistic axial jets of energy of order
$10^{51}$ ergs originating within the collapsed core can
initiate a bi-polar asymmetric supernova explosion that is
consistent with the spectropolarimetry (Khokhlov et al. 1999;
Khokhlov \& H\"oflich 2001; H\"oflich, Khokhlov \& Wang 2001).
Some imbalance in axial jets can also account for pulsar
runaway velocities, specifically velocities that are parallel to the
spin axis (Helfand, Gotthelf \& Halpern 2001, and references
therein).  While a combination of neutrino-induced and jet-induced
explosion may prove necessary for complete understanding of
core-collapse explosions, jets as computed by Khokhlov et al. are
sufficient. In this paper, we will ignore the possibility of
neutrino-induced explosions as we focus on the possibilities of jet
formation and jet-induced supernovae. For purposes of discussion, we
will consider the sort of structure that forms a stalled shock after
core bounce.

Further work on jet-induced supernovae has been done by
Khokhlov \& H\"oflich (2001) and H\"oflich, Khokhlov \& Wang (2001).
These papers present well-resolved, three-dimensional jets that propagate
from within the region of the original iron core until breakout
from the star or stoppage in a red giant envelope
in a single calculation that does not
require artificial halting, rescaling, and restarting the jet.
A bow shock forms at the head of the jet and spreads roughly cylindrically
around each jet.  The stellar matter is shocked by the bow shock and
acts as a high-pressure confining medium by forming a cocoon around
the jet.  The jets become long bullets of high-density material moving
through the background low-density material almost ballistically.
Spreading is limited by a secondary shock that forms around each jet
between the jet and the material already shocked by the bow shock.
The laterally expanding bow shocks generated by the jets move
towards the equator where they collide with each other.
The result is that the material in the equatorial plane is
compressed and accelerated more than  material in other directions
(excluding the jet material).  The result is that heavy elements
(e.g. O, Ca) are characteristically ejected in tori along the equator.
Iron, silicon and other heavy elements in Cas A are distributed
in this way (Hwang et al. 2000), and there is some evidence for this
distribution in SN 1987A (Wang et al.  2001b).

The presence of a red giant envelope can
stop the jet and dissipate the effects of the asymmetry in the
outer envelope, but the asymmetry will still be manifest in the core.
Radioactive matter ejected in the jets can alter the ionization
structure and hence the shape of the photosphere of the envelope even
if the density structure is spherically symmetric (H\"oflich, Khokhlov
\& and Wang 2001).  This will generate a finite polarization, even
though the density distribution is spherical and the jets are stopped
deep within the star and may account for the early
polarization observed in Type II supernovae (Leonard et al. 2000; Wang
et al. 2001).  The jets are unstable to Kelvin-Helmholtz
instabilities, but the growth time is long compared to the jet
propagation time and no  significant modulation of the jet is
observed.  

Faster jets tend to make weaker explosions by propagating so
rapidly and narrowly through the star
that there is relatively little time to generate the transverse
shocks that cause the explosion.  The opening half angle of the
jet is approximately (Wheeler et al. 2000),
\begin{equation}
\label{jetangle}
\theta \simeq \frac{v_{env}}{v_{bow}}\simeq 0.1~{\rm rad}
\frac{v_{env,8}}{v_{bow,9}}\simeq 5^o\frac{v_{env,8}}{v_{bow,9}}.
\end{equation}
where $v_{env}$ is the speed of sound of the envelope and
$v_{bow}$ is the speed of the bow shock, which is less than
that of the inflow velocity of the jet material.  As $v_{bow}$
approaches the speed of light, the jet will be very narrow and
affect a very small volume of the star as it propagates out.
This feature may be relevant to
the outcome of the explosion process and to the possibility of
making \grbs, as we will discuss in \S 4.3.
Radiative transfer effects within optically thin portions of
the jet may lead to significant instabilities in faster,
lower density, jets (H\"oflich, private communication).
For related work in the context of black hole formation, see
MacFadyen \& Woosley 1999; Aloy, et al. 2000; MacFadyen, Woosley
\& Heger 2000; Zhang \& Woosley 2001).

The question then arises as to the mechanism of the production
of the jets in routine core collapse events.  A significant role for
rotation and magnetic fields is an obvious candidate.  Any
mechanism that purports to account for routine pulsar
formation may involve large transient magnetic fields, but must
ultimately be consistent with the distribution of deduced
dipole strengths of ``normal'' pulsars of $\sim
10^{12}$--$10^{13}$ G and with the distribution of initial rotation
periods.  There is growing evidence for a considerable spread
in these quantities ({\it e. g.} Kaspi, et al. 2001 and references therein).
In a minority of cases, the final dipole
field might be consistent with a value of $\sim 10^{15}$ G,
yielding a ``magnetar'' (Duncan \& Thompson 1992).
Growing evidence for neutron
stars of that field strength has been obtained by RXTE and other
facilities (Kouveliotou et al. 1999; Ibrahim et al. 2001).
The nature of the birth event of a magnetar is a separate
problem that is significant in its own right.  The ensemble
of normal pulsar and magnetar births must also be consistent
with observed nucleosynthetic abundances, a potentially
crucial constraint on jet-induced supernova models
that we discuss in \S 4.

Possible physical mechanisms for inducing axial jets,
asymmetric supernovae, and related phenomena driven by
magneto-rotational effects were considered by
Wheeler et al. (2000).  The means of
amplifying magnetic fields by linear wrapping associated with
differential rotation in the neutron star and possibly by
$\alpha$~--~$\Omega$ dynamos was discussed.  Attention was
focused on the effect of the resulting
net dipole field on the torquing
of the infalling plasma and newly formed neutron star and on the
creation of strong Poynting flux (in the form, initially, of
ultrarelativistic MHD waves) at
the time-variable speed of light circle in analogy to pulsar
radiation mechanisms, albeit buried deeply in the core-collapse
supernova ambience.  Allusion was made to the existence and
possible role of the primary toroidal field that is expected to
form in routine collapse where protoneutron star rotation
rates are not expected to trigger $\alpha$~--~$\Omega$ dynamos
and exponential field growth.

In this paper, we will explore the capacity of the predominantly toroidal
field 
generated near the core-mantle boundary to
directly generate axial jets within core-collapse
conditions by analogy with magneto-centrifugal models of jets
in AGN (Koide, Shibata \& Kudoh 1997;
Meier et al. 1997; Romanova et al. 1998; Ustyugova et al.  1999;
Meier, 1999; Koide, Meier, Shibata \& Kudoh 2000; and references therein).
The magneto-centrifugal models in the literature
usually focus on black hole conditions where the field is
anchored in the disk and at infinity.  The physical picture
discussed in this paper may prove especially robust since the
field is anchored in substantial, corporeal objects:  the
outer, still collapsing layers of the progenitor star and the
collapsed object --- a neutron star.  We find that the
production of a strong toroidal
field, substantially stronger than the $10^{12}$ G dipole field of a
pulsar, is nearly inevitable, and strong axial jets driven by
that field equally so.  The mechanisms described here also may
prove to have a close astrophysical analog in the
production of supersonic magneto-centrifugal jets in young
stellar objects where also there is an outer envelope of
infalling matter and an inner, differentially rotating object
(Konigl \& Pudritz 2000; Lery \& Frank 2000; Ouyed, Pudritz
\& Stone 1997).

Our working hypothesis is thus that jets leading to asymmetric
supernova explosions are associated with routine core collapse
and pulsar formation in roughly 90\% of the cases.
Perhaps 10\% of the core collapse events would be associated
with magnetars.  Magnetar formation might be associated with
exceptionally strong explosions and asymmetries, but this
is not necessarily the case.  An even smaller fraction,
perhaps $10^{-3}$ to $10^{-4}$ (Scalo \& Wheeler 2002a) might involve
black hole formation and the production of \grbs.

In \S 2 we outline the circumstances by which a newly formed
neutron star could 
yield MHD jets.  In \S 3, we discuss the basic physics of MHD jets
and their formation.   Section 4 explores
the application of the physical principles of jet formation to
supernova conditions, how magneto-centrifugal jets could be
generated in core collapse, and some of their expected properties.
We sketch a scenario in which extreme values of the rotation
and magnetic field could lead to the failure of the initial supernova
explosion with the subsequent collapse of the core to form a
black hole and the attendant possibility to form a \grb.
In \S 5 we summarize our conclusions.

\section{The Production of Jets from Neutron Star Formation}

To understand routine, pulsar-forming, strongly asymmetric supernovae,
there must be a deeper understanding of the generation of jets
during the process of core collapse to form neutron stars.

We identify three stages where magneto-rotational processes
may come into play in the collapse process:

\noindent
(1) The initial collapse or LeBlanc-Wilson phase,
hereafter referred to as the ``LeBlanc-Wilson jet'' or ``LW jet''
(LeBlanc \& Wilson 1970).  In this phase, the
wrapping of field lines of the original, pre-collapse core
field {\em within} the nascent neutron star can give rise to a
magneto-centrifugal
jet or rising magnetic ``bubble'' that forms along the rotation
axis at $R\sim5\times10^7$ cm and $\rho \sim 10^8$ g cm$^{-3}$.
Parameters typical of modern pre-supernova calculations imply
that the LeBlanc-Wilson stage is probably too weak to form a strong
jet and explosion with observed supernova energies.
Constraints from the production of {\it r}-process material also argue
that this phase must not be robust (\S 4.1).
 
\noindent
(2) The protopulsar phase.  There will be further
wrapping of field lines by linear amplification
of the contracting, cooling, de-leptonizing protoneutron
star that is differentially rotating with respect to the
infalling matter and the uncollapsed outer portions of the star.
The toroidal fields can form within the newly formed, contracting,
neutron star, at the shearing boundary between the neutron star
and the infalling matter, and within the infalling matter.
Between $10^{51}$ and $10^{52}$ ergs of rotational energy are
available during the contracting deleptonization phase.  A
substantial portion of this energy will go into a toroidal field
generated by differential rotation with a strength
$\sim$~$10^{14}$~--~$10^{16}$ G.
The jets are launched by toroidal magnetic ``springs'' that
evolve into a steady-state, Blandford-Payne MHD wind (\S 3.1),
the energy and momentum of which eject the stellar envelope.
This is essentially
the earliest phase in the birth of what will be the pulsar
remnant, and is characterized by a period of rapid spin-down
due to the production of the bipolar MHD outflow.
This is a natural environment
for a magneto-centrifugal jet and may be the routine
mechanism for jet-induced, asymmetric supernova explosions.
We will refer to this as the ``protopulsar jet'' or
``slow toroidal jet'' phase.

\noindent
(3) In relatively rare circumstances, depending on the
distribution of rotation rates of the pre-supernova cores, the
new neutron star may attain a rotation period of $\sim 1$ msec
after deleptonization where
the shear can compete with the convective
motions and an $\alpha$~--$\Omega$ dynamo becomes possible.
The resulting exponential growth of radial and toroidal fields
generates fields of strength $10^{16}$ G or more.
The rapid growth of the field
coupled with the large rotational energy reservoir associated
with millisecond rotation could lead to especially large
explosions.  A mix of a
magneto-centrifugal jet and the emission of a strong Poynting
flux at the light cylinder could lead to power of
$\sim 10^{52}~\rm{erg~s^{-1}}$ in extreme cases.  It is
possible that such cases have been witnessed as
``hypernovae,'' but strong asymmetries can
affect estimates of energetics (H\"oflich, Wheeler \& Wang 1999).
We will return to this topic in the conclusions.  Another
alternative is that very extreme conditions lead to especially
fast, narrow jets that actually put {\it less} energy into the
star (Khokhlov \& H\"oflich 2001).
We will refer to these as ``fast toroidal jets.'' In this case,
the MHD explosion mechanism may {\em fail} in the first instance
and lead to the formation of a black hole.  The formation
of a rapidly-accreting black hole may, in turn, lead to the formation
of  an even faster, highly relativistic
jet that interacts with the first, slower jet to
produce a \grb.  We will explore this possibility in \S 4.3.

Wheeler et al. (2000) discussed the possible
physical processes, associated
timescales, and energetics that could lead to the production of pulsars,
jets, and asymmetric supernovae and to a $10^{15}$ G magnetar
in more extreme circumstances in the collapse of a massive stellar core.
Wheeler et al. (2000) noted that
when the neutron star contracts
and speeds up two significant things may happen.  One is that
the rotational energy increases.  The energy becomes significantly
larger than that required to produce a supernova and, especially for
recent estimates (Frail et al. 2001; Kumar \& Panaitescu 2001),
to drive a cosmic \grb.  In addition, the light cylinder
may contract from a radius large compared to the Alfv\'en radius
to a radius comparable to that of the neutron star.  This could
disrupt the structure of an organized dipole field and promote
the generation of intense MHD waves.  The frequency of the MHD
waves, $\sim\Omega_{NS}
\sim10^4$ Hz, would always be less than the plasma frequency, so these
waves would be reflected internally and perhaps also channeled
up the rotation axis.  Collimation of the flow of
energy may thus be expected.

A jet emerging from deep in the collapsing iron core
of a star will emerge from the helium core in 5 to 10
seconds.   Wheeler et al. (2000; see also Nakamura 1999)
argued that a jet launched after the deleptonization
contraction of the neutron star
could arrive at the surface of the helium star at about the
same time.  This argument is incorrect.
The 5 to 10 seconds associated with the loss of neutrinos
in the deleptonization phase represents the characteristic
time scale for the luminosity, but the luminosity is
enhanced at later phases by the contraction which results
in a higher temperature at the neutrinosphere.  The time scale
for the contraction of the radius and hence the spin up of
the neutron star is actually considerably shorter, of
order one second (Wilson \& Mayle 1993).  This means
that any stronger jet launched during the contraction will
catch up to and dominate any magnetic ``bubble'' or jet
launched promptly during the initial collapse before the
first jet can emerge from the helium core.  If this
results in a supernova explosion, then any {\it r}-process
material ejected, however slowly,
from the neutron star in the first phase will be
at least partially ejected in the explosion.  Furthermore,
if the collapse results in the formation of a black hole with the
generation of a third, relativistic, jet, this jet will, in turn,
catch up with any jet matter ejected earlier.  The time when this
happens will depend on the delay between the first iron-core collapse
and the second protoneutron star collapse to form the black hole (\S 4.3).

\section{Magneto-centrifugal Jets}

\subsection{The Blandford-Payne Mechanism:  ``Spring'' and ``Fling''}

The steady-state theory of magneto-rotationally-driven outflows was
discussed by Blandford and Payne (1982) in the context of magnetized
disks. Numerical treatments of such
collimating winds have been performed by Ustyugova  et al.
(1995), Ouyed, Pudritz \& Stone (1997), Krasnopolsky, Li \& Blandford (1999)
and others.  In the steady-state model, a magnetic field anchored
at the base in a rotating object transfers energy and momentum
to the plasma frozen onto the field lines.  Differential rotation
between the inner and outer regions of the outflowing plasma causes
the field lines to be swept backward in an open helix; 
and the overall rotation of this structure pushes plasma upward parallel
to the rotation axis and slowly compresses it toward that axis.
Eventually this creates a collimated bi-polar outflow at large distances
from the central object.
This process is sometimes referred to as the ``fling'' model, since it
employs centrifugal action to initially accelerate plasma along field
lines.  The forces that eventually complete the acceleration
and collimation are magnetic pressure upward parallel to the axis
and hoop stress.  

A similar, but time-dependent, process was investigated by Uchida \&
Shibata (1985) and more recently by Kudoh, Matsumoto \& Shibata (1998),
Uchida et al. (1999), and Koide  et al. (2000).  In this
case an initially {\em poloidal} magnetic field ({\it e.g.}, uniform or
dipole) threads a differentially rotating medium, such as a thick disk
or torus.  As the evolution
proceeds, the differential rotation coils the field in a barber-pole
fashion, lifting and pinching material along the rotation axis. 
Depending on the strength of the field and the differential
rotation, the early evolution of this system can be quite dynamic.
The field will not have had time to expand to a broad open structure
and approach the Blandford-Payne steady state. Instead, it may wind
quickly to very large toroidal strengths and then expand vertically
along the rotation axis, dynamically lifting and pinching plasma
along the way.  For this reason, this more dynamical process is sometimes
referred to as the ``spring'' mechanism for jet outflow.  Both
spring and fling, however, use the same magnetic pressure and hoop
stresses to perform the final acceleration and collimation.
While both processes were first discussed in the context of accretion
disks, they will have close analogies in collapsing supernova cores,
especially if the core rotation results in significant core flattening
and an accretion-torus-like structure surrounding the core.

For non-relativistic outflow, the power output of an MHD flow is
given by (Blandford \& Payne 1982)
\begin{equation}
\label{L-MHD}
L_{MHD} \; = \; B^2 \, R^3 \, \Omega \, / \, 2,
\end{equation}
where $B$, $\Omega$, and $R$ are, respectively, the magnetic field
at the base of the jet, the jet angular velocity, and the size scale of
the region where outflowing matter is injected along the field lines.

\subsection{The Magnetic Switch}

If, in the above cases, the magnetic field is sufficiently strong
or the rotation sufficiently fast, then the power output can exceed a
critical dynamical value, given by the ratio of the escape energy
to the free-fall time. For the case of a constant density corona of
radius R, this can be expressed as:
\begin{equation}
\label{L-crit.rot}
L_{crit} \; = \; \frac{E_{esc}}{\tau_{ff}} \; = \; \frac{m_{ej}}{R} \, 
\left( \frac{GM}{R} \right)^{3/2},
\end{equation}
where the escape energy is taken to be $GM m_{ej}/R$.
Simulations of magnetized accretion disk coronae with a 
height comparable to the radial extent ($m_{ej} \sim
4\pi R^3 \rho$; see Meier et al. 1997; Meier 1999) have shown that
under these conditions much of the MHD power may be converted
quickly into kinetic energy, creating a fast, highly-collimated
jet close to the injection region, and that this condition can
persist well into the steady state.  In the current situation, 
the mass to be ejected that determine the escape energy is that in a cone
of fractional solid angle $f_{\Delta\Omega} = \Delta\Omega/4\pi$,
where $\Delta\Omega$ is the solid angle of the jets emerging
from the poles of the neutron star.  Assuming that the mass in this
cone is concentrated near the lower portions where the density is
$\rho$ at radius $R$, we take $m_{ej} \sim 4/3\pi R^3 f_{\Delta\Omega}$ and
hence:
\begin{equation}
\label{L-crit.ns}
L_{crit} \; = \; \frac{E_{esc}}{\tau_{ff}} \; = \; \frac{4 \pi}{3} \, \rho
\, R^2 \, f_{\Delta\Omega} \, \left( \frac{GM}{R} \right)^{3/2}.
\end{equation}
For $L_{MHD} < L_{crit}$,
the outflow will be a classical Blandford-Payne-type outflow,
collimating slowly with the jet thrust spread over a broad region.
(See, {\it e.g.}, the steady-state simulations of Krasnopolsky
et al. 1999).  This criterion for a slow jet
can be expressed in terms of
a dimensionless parameter, $\nu$, as
\begin{equation}
\label{nu-crit}
\nu \; \equiv \; \left( \frac{L_{MHD}}{L_{crit}} \right)^{1/2}
\; = \; \frac{v_{A}}{v_{esc}} \, f_{\Delta\Omega}^{-1/2} \, 
\left( \frac{3 \Omega}{\Omega_{K}}
\right)^{1/2} \; < \; 1.
\end{equation}
For $\nu > 1$, the jet may have a much higher velocity
(with $v_{jet} \sim v_A > v_{esc}$) early on,
with the jet thrust concentrated in a very
narrow axial spindle (Meier et al. 1997).

\section{Magneto-Centrifugal Jets in Supernovae}

We will now explore the application of the ``spring" and
``fling" and fast and slow jets to the supernova collapse
problem and consider in more detail the three stages of the
collapse that may lead to MHD jets as outlined in \S 2.

\subsection{The LeBlanc-Wilson Jet}

The first type of MHD outflow we consider is that discovered
by LeBlanc \& Wilson (1970; hereafter LW), also treated later by Symbalisty
(1984). LW studied the magneto-rotational collapse of a 7 \m\
iron core by numerically solving the two-dimensional MHD
equations coupled to the equation for neutrino transport.
Their simulations showed that a ``circulation vortex'' formed
along the rotation axis, at about the time of core bounce, amplified
the magnetic field there to of order $\sim 10^{15}$ Gauss, and
created two oppositely-directed,  magnetically-driven, high-density,
supersonic jets of material  emanating from inside the protoneutron
star.  In the calculation of  LW, the jet formed at about 50 km, in
that of Symbalisty, at about 30 km.  LW estimated that their
jet carried away $\sim 10^{32}$ g with $\sim 1 - 2 \times 10^{51}$
erg in $\sim 1$ s.  Khokhlov {\it et al.} (1999) assumed
characteristics of the LeBlanc-Wilson jet for their proof of
principle calculations.

The calculation of LW was extended by Meier et al.
(1976) to treat stellar cores with mass, central density, and rotational
characteristics more commensurate with the results of modern advanced
stellar evolutionary calculations.  One of their main constraints on
the LW mechanism was that it not over-produce {\it r}-process material
in the Galaxy, a limit of roughly $2 \times 10^{-4}$ \m\
of highly-neutronized ejecta per supernova.  Meier et al. pointed
out that this constraint was not compatible with the notion
that the LW mechanism
provides the explosion energy for most supernovae: if it were to do so,
then the {\it r}-process elements would be over-produced by a factor
of $\sim 1000$.  Therefore, either the LW jet provides the explosion
energy (and most of the Galaxy's {\it r}-process material) in only a very
few rare supernovae, {\em or} LW jets or bubbles are quite common in
most core-collapse supernovae, but eject only $\sim 10^{-4}$ \m\
per event, and are not important in the supernova's overall
energetics because the progenitor did not have especially large
initial rotation and magnetic field.

In the latter, more plausible, case, Meier  et al. (1976) showed that
the  LW events would proceed as ``buoyancy instabilities,'' with the
magnetized  material rising subsonically up the rotation axis of the
protoneutron  star and bursting into the mantle above.  In the
more modern language of MHD winds and jets, these LW events will
result in Blandford-Payne-type  MHD wind flows emanating from the
upper interior of the protoneutron star.  These  MHD ``fountains'' will be
initially sub-sonic, relative to the core sound  speed, and rather
confined by the high pressure, but when they  burst into the mantle,
they may be supersonic with respect to the sound  speed there.  In a
completely-bare core collapse, with little or no mantle  to disrupt
these weak outflows, these events eventually could accelerate  to
supersonic speeds and form jets of {\it r}-process material, but in
most supernovae the LW ejecta are likely to quickly mix with the
polar  regions of the infalling mantle material.

Taking the collapsing protoneutron star to be about 1.5 \m\ in mass, with
initial and final radii of $\sim 10^{8}$ cm and $\sim 5 \times 10^{6}$ cm,
respectively, we estimate from Meier et al. (1976) that the magnetic
field  still will be rather high (few $\times 10^{15}$ Gauss) in the
circulation-vortex region near the axis, but that the {\em differential}
rotation rate there will be rather low, $<$ 100 rad s$^{-1}$.
From the {\it r}-process constraints and equation (\ref{nu-crit}) we find
that the magnetic switch parameter $\nu$ will be of order 0.03 or less
(for $f_{\Delta\Omega}\sim1$),
nearly independent of the exact values of the collapse parameter and the
efficiency of conversion of rotational energy to magnetic energy.  This
result that the MHD outflow is well below the magnetic switch point
is consistent with the LW fountain being an initially subsonic,
Blandford-Payne-like, polar outflow into the mantle.

\subsection{The Slow Toroidal Jet}

\subsubsection{Generation of the Toroidal Field}

Rotation alone can inhibit the contraction of the newly-formed neutron star
(Symbalisty 1984; M\"onchmeyer, et al. 1991;
Fryer \& Heger 2000).  However the addition of magnetic
fields provides a source of dissipation that will
allow further contraction, thus enhancing the rotational
energy, the differential rotation, and ultimately the field energy.
The presence of even a small field, therefore, can significantly
increase the possibility of creating strong axial jets.
As discussed in the introduction, Wheeler et al. (2000)
considered the linear wrapping phase of field amplification
due to differential rotation within the neutron star
(Meier, et al. 1976; Klu\'zniak \& Ruderman 1998;
Ruderman, Tao \& Klu\'zniak 2000), but they did not emphasize the
resulting toroidal field, only the dipole component
of that field (taken to be $\sim 1 \%$ of the toroidal field)
and its possible effects in concert with
the Alfv\'en surface and speed of light circle.
Here we argue that the principle focus should, rather, be
on the strong toroidal field itself.  By analogy to
the work on the creation of jets by MHD processes in accretion
disks (\S 3) the toroidal field can act as a ``spring'' to launch jets.
We expect this sort of jet to overtake and supplant the original
LeBlanc-Wilson jet or bubble, as illustrated schematically in
Figure 1.

The original rotational energy of the protoneutron star (PNS) is approximately,
\begin{equation}
\label{littleErot}
E_{\rm rot,PNS}\simeq\frac{1}{2}I_{PNS}\Omega_{PNS}^2
\simeq 9\times10^{50}\ {\rm erg}
\left(\frac{M_{NS}}{1.5 {\mm}}\right)
\left(\frac{P_{PNS}}{25~{\rm ms}}\right)^{-2}
\left(\frac{R_{PNS}}{50~{\rm km}}\right)^2,
\end{equation}
and the rotational energy after the deleptonization and contraction is about,
\begin{equation}
\label{ErotNS}
E_{rot,NS}\simeq\frac{1}{2}I_{NS}\Omega_{NS}^2
\simeq2\times10^{52}~{\rm erg}
\left(\frac{M_{NS}}{1.5 {\mm}}\right)
\left(\frac{P_{NS}}{1~{\rm ms}}\right)^{-2}
\left(\frac{R_{NS}}{10~{\rm km}}\right)^2.
\end{equation}
It is not clear how much of the latter energy can be dissipated
in the magnetic field, but there is substantial energy for
a supernova for any final rotation period of less than about 7 msec.
Strictly speaking only the energy associated with the
differential rotation can be tapped, but the neutron
star is always in strong differential rotation compared to
the progenitor star.  There
may be constant torques on the neutron star until the
explosion succeeds or fails, so this issue is complex
and involves the magnetic fields we will now explore.

If the pre-collapse progenitor core has a field strength
comparable to that of a magnetized white dwarf,
$\sim10^8$ G, then a field of $\sim10^{12}$ G
could arise during the collapse simply from flux-freezing.  This
field can be amplified further
by differential rotation in the neutron star (Meier et al. 1976;
Klu\'zniak \& Ruderman 1998) to produce a strong toroidal field.
For a collapsing white dwarf, The protoneutron star is likely
to be differentially rotating with angular velocity
$\Omega$ increasing outward since the protoneutron
star will be ``stiffer'' than the
highly degenerate white dwarf progenitor (Ruderman,
Tao \& Klu\'zniak 2000).   For a neutron star formed by
homologous iron core collapse, the angular velocity 
of the homologous core will increase
outward within the homologous core at core bounce, but after 
bounce the angular velocity
is nearly constant in the inner core and decrease with radius beyond
the boundary of the initial homologous core (Akiyama, Wheeler \& 
Meier 2002).  The process of field generation and expulsion
will depend on the details of the differential rotation
(Akiyama et al. 2002; Akiyama, Wheeler \& Lichtenstadt 2002).

For this ``wrapping'' mechanism, the field will grow linearly
with time and be proportional to the seed field.
After $n_\phi$ revolutions of the neutron star,
the initial seed (poloidal) field will produce a toroidal field,
\begin{equation}
\label{Bphi}
B_\phi\simeq2\pi n_\phi B_p,
\end{equation}
where $B_p$ is the initial seed poloidal field.
The field will be limited by buoyancy, which will operate to expel
the field from the site where it is generated. The field at this buoyancy
limit, $B_b$, is given by,
\begin{equation}
\label{buoyancy}
\frac{B_b^2}{8\pi}\gta f_{b} \rho c_s^2,
\end{equation}
where $f_{b}$ is the fractional difference in density
between the rising flux tube elements and the stellar material.

There are two regions where there is likely to be large shear
and hence field growth.  One is within the neutron star.  The
angular velocity can be large there, but, as remarked above, the
gradient is rather flat after core bounce.  A complication is that the
dynamics of the LW phase may already have amplified the field within
the  neutron star so the seed field for the subsequent wrapping phase
may be larger.  The other region where strong field amplification is
expected is at the boundary of the protoneutron star where the angular
velocity is less than in the neutron star, but the difference in the
angular velocity with respect to the postshock,  settling matter
results in a rather large gradient.

After bounce, the central density of the neutron star is
about $2\times10^{14}$ g cm $^{-3}$.  The boundary of the homologous
core at R $\sim 10$ km is initially at about $10^{13}$ g cm $^{-3}$\
at bounce, but settles to somewhat higher densities,
$\sim 3\times10^{13}$ g cm $^{-3}$, after some relaxation
(Burrows, Hayes \& Fryxell 1995; Janka \& M\"uller, 1996;
Mezzacappa et al. 1998).
For sound speed $c_s \sim 10^{10}$ \cms, density
$\rho\sim10^{14}$ g cm $^{-3}$ and $f_{b}\sim0.01$ the
critical value of the field at the onset of buoyancy is,
\begin{equation}
\label{Bbuoyancy}
B_b\simeq6\times10^{16}~{\rm G}~f_{b-2}^{1/2}\rho_{14}^{1/2},
\end{equation}
where $f_{b-2} = f_b/0.01$ and $\rho_{14}=\rho~/10^{14}$ g cm $^{-3}$.
The number of revolutions to reach the buoyancy limit
$B_b$ is,
\begin{equation}
\label{nwrap}
n_f \simeq \frac{B_b}{B_0}\frac{1}{2\pi}
\simeq10,000~B_{0,12}^{-1}~f_{b-2}^{1/2}\rho_{14}^{1/2}.
\end{equation}
For $B_0\sim10^{12}~{\rm G}$ and $\rho \sim 10^{14}$ g cm $^{-3}$,
the linear amplification time scale before the
field is expelled by buoyancy is thus,
\begin{equation}
\label{twrap}
t_f\simeq n_fP\sim 10~{\rm s}~P_{ms}
~B_{0,12}^{-1}~f_{b-2}^{1/2}\rho_{14}^{1/2},
\end{equation}
where $P_{ms}$ is the rotational period in msec.

The second region, around the boundary of the protoneutron star at
R $\sim 50$ km, has a typical density $\sim 10^{11}$
g cm $^{-3}$ after the shock stalls at about 200 km.
For sound speed $c_s \sim 3\times10^{9}$ \cms, density
$\rho\sim10^{11}$ g cm $^{-3}$ and $f_{b}\sim0.01$, the
critical value of the field at the onset of buoyancy at the
protoneutron star boundary is,
\begin{equation}
\label{Bbuoyancy2}
B_b\simeq5\times10^{14}~{\rm G}~f_{b-2}^{1/2}\rho_{11}^{1/2},
\end{equation}
the number of revolutions to reach the buoyancy limit is,
\begin{equation}
\label{nwrap2}
n_f \simeq 80~B_{0,12}^{-1}~f_{b-2}^{1/2}\rho_{11}^{1/2},
\end{equation}
and the linear amplification time scale before the
field is expelled by buoyancy is,
\begin{equation}
\label{twrap2}
t_f \sim0.08~{\rm s}~P_{ms}
~B_{0,12}^{-1}~f_{b-2}^{1/2}\rho_{11}^{1/2}.
\end{equation}

Linear wrapping, therefore, is slow at first, but is likely to
accelerate with the contraction and spin up of the neutron star.
Because there will be a gradient in density and sound speed
from within the neutron star to the post-shock, settling material,
there will be variations in the time at which the field reaches
the buoyancy limit and the value of the field at that limit around
the surface of the neutron star.   The magnetic field will be subject
to wrapping by differential rotation throughout the collapsing medium
since the field is ultimately anchored in the outer, slowly rotating
progenitor.   

In general, the epoch of maximum expulsion of the toroidal magnetic
field is seconds for the strong field within the neutron star to
a fraction of a second somewhat beyond the neutron star, a range that
spans the cooling, contraction time of $\sim 1$ s.
An important point is that this process of wrapping and expelling
field does not necessarily halt after one characteristic time.
The toroidal field will build up to the buoyancy limit and then
begin to be expelled.  Field of this strength, $\gta 10^{16}$ G
near the homologous core boundary, $\gta 10^{14}$ G 
at the boundary of the protoneutron star, 
will continue to pump out as the differential
rotational energy is dissipated with a time scale that is much
greater than the characteristic time of the differential rotation
$\sim \Omega^{-1}$.

This discussion illustrates that the buoyancy time and
subsequent, longer term dissipation of
the rotational energy into magnetic field can be
short enough to be a significant effect on the radial
contraction time of the neutron star.
There will be a gradient of buoyancy field that breaks
out of the neutron star as a function of time and location,
the nature of which needs to be explored.
The characteristic timescales for
growth and buoyancy throughout the neutron star are
long compared to dynamical times and hence all the associated
MHD processes will be quasi-static on that
timescale.  The energy in the magnetic field therefore does
not directly power the explosion; that is, the magnetic pressure
does not build up to explosive values and unbind the star.  Rather,
the magnetic field serves as a conduit to transmit
the energy of differential rotation into directed jet energy
and momentum over many dynamical times, and it is the jet
dynamical power that creates the explosion.

The rotational amplification of the field and
its expulsion will go in the direction of eliminating
the shear, but the infall of more slowly rotating
matter will always maintain a differential rotation with negative
gradient in $\Omega$ at the neutron star boundary.
The creation and expulsion of field from this layer
should continue until the explosion of the star
cuts off the infall.  Initially, the magnetic field
should have negligible feedback on the dynamics;
but, as the magnetic field builds up, it will affect the inflow.

\subsubsection{Spring and Fling in Supernovae}

We can now assess the affect of the toroidal field on the
physics of jet formation in the supernova ambience.
The Blandford-Payne power (eqn. [\ref{L-MHD}]) can
be expressed as:
\begin{equation}
\label{L-MHDSN}
L_{MHD}\simeq\frac{B^2R^3\Omega}{2}
\simeq3\times10^{52}~{\rm erg~s^{-1}}B_{16}^2 R_{NS,6}^3
\left(\frac{P_{NS}}{10~{\rm ms}}\right)^{-1},
\end{equation}
for parameters representative of the protoneutron star interior.
Note that the effect of the smaller field near the protoneutron star 
boundary, $\sim 10^{15}$ G, would be substantially offset by the
larger radius, $\sim 50$ km, so that the power is roughly the
same in both key locations.  If the neutron star can sustain a breakout
buoyant field of $10^{16}$ G from deep within,
or a field of $10^{15}$ G near the protoneutron star
boundary, for $\gta$ 0.1 s as the
differential rotational energy is tapped, a strong supernova could
result.

For the conditions of interest here, the critical parameter
of the magnetic switch mechanism (eqn. [\ref{nu-crit}])  can be written as:
\begin{equation}
\label{nu-SN}
\nu\simeq 0.05~B_{15} \left(\rho_{11} P_{ms} f_{\Delta\Omega}\right)^{-1/2} 
R_6^{5/4} M_{NS}^{-3/4},
\end{equation}
where for convenience here and below, the mass of the neutron
star has been normalized to 1.5 \m. This equation
says that there is a locus of critical values at which the
switch would be activated and an especially high speed
jet created, defined by $\nu$=1, namely; 
\begin{equation}
\label{crit-locus}
B_{crit}\simeq 2\times10^{16}~{\rm G}~
\left(\rho_{11} P_{ms}f_{\Delta\Omega}\right)^{1/2} R_6^{-5/4} M_{NS}^{3/4}.
\end{equation}
The value $\nu = 1$ corresponds to a (critical) MHD power of
(eqn. [\ref{L-crit.ns}]),
\begin{equation}
\label{L-critNS}
L_{crit}\simeq1.2\times10^{54}~{\rm erg s^{-1}}\rho_{11} f_{\Delta\Omega}
 R_6^{1/2} M_{NS}^{3/2}.
\end{equation} 
The conditions to make this switch in jet character could
be found in collapse conditions for especially rapid rotation
and hence large field strength or if a strong field becomes
buoyant and ``floats" into lower density conditions,
or as the inflow density declines.  Note the potentially important
role of the collimation of the jet as reflected in the parameter
$f_{\Delta\Omega}$.  This parameter would be of order unity for
a slow Blandford-Payne jet with a broad base and size of order
the diameter of the neutron star in which case a rather large
luminosity is required to trigger the fast jet.  If, however,
the jet is tightly collimated as numerical simulations show for
the case with ``switch on" or as indicated by \grb\ jet collimation,
then $f_{\Delta\Omega}$ could be quite small $\lta 10^{-2}$ and
a significantly more modest luminosity would be adequate to trigger 
the switch and the fast jet.  Hoop stresses in the field surrounding
the neutron star could lead to such tight collimation even for 
the slow jet, but this requires more study.  

The speed of the jet when the switch is ``on" will be somewhere
between the Alfv\'en speed
\begin{equation}
\label{alfvenspeed}
v_A \simeq \left(\frac{B^2}{4\pi\rho}\right)^{1/2} \simeq
9 \times 10^{8}~{\rm cm~s^{-1}}B_{15}\rho_{11}^{-1/2},
\end{equation}
and the ``equipartition" speed (where $L_{MHD} \simeq L_{kinetic}$)
\begin{equation}
\label{equipspeed}
v_E \simeq \left( v_A^2 \, \Omega R \right)^{1/3} \simeq
1.7 \times 10^{9} ~{\rm cm~s^{-1}} B_{15}^{2/3} \rho_{11}^{-1/3}
P_{ms}^{-1/3}\hspace{1mm} ,
\end{equation}
which is $\sim v_A$ if, as we expect, the field amplification saturates
and $v_A \sim \Omega R$.  These numbers suggest the possibility
of mildly relativistic flows if B exceeds $10^{16}$ G, 
with the outflow speeds depending
critically on the maximum Alfv\'en speed attained in the region
outside the rapidly rotating core.
These aspects clearly need further study.
 
These order of magnitude estimates serve to illustrate
that this spring mechanism might be quite common in core collapse,
generating bipolar outflows of sufficient energy and velocity that
they could be the sole means of supernova explosion.  At minimum,
the presence of such a jet would distort an already successful
explosion, thereby creating the observed asymmetries.

\subsubsection{Dynamics and Nucleosynthesis}

There is a danger that the process we have outlined here is too effective
in one regard, the production of neutron-rich elements.  We have noted
that this is a severe constraint on the LeBlanc-Wilson jet, such that the
LW phase cannot be responsible for routine supernovae without severely
over-producing neutron-rich species (\S 4.1).  If conditions are just right,
a jet-like explosion might account for the {\it r}-process
(Symbalisty, Schramm \& Wilson 1985; Cameron 2001).
For an explosion based on the protopulsar jets, we have to seek
a balance between an effective deposition of energy, namely a jet
that is not too fast or narrow, and one that is not too broad that
will lift off too much neutron-rich matter from deep in the collapse
environment. Resolution of this dilemma will require more quantitative
analysis, but here we sketch some of the basic issues.

There is a major advantage to a jet that forms near or beyond the
protoneutron star boundary where the neutronization will not
be as extreme as it is deep within the protoneutron star where
the LW jet forms and where large toroidal fields will form
in the subsequent wrapping, contraction phase.  Nevertheless,
the infalling matter will still be neutron rich (the electron
fraction Y$_e$ plummets from $\lta 0.5$ at the standing shock to 
$\gta 0.2$ for R $\gta 50$ km; e.g. Mezzacappa et al. 1998), 
and excessive ejection would
again violate limits on neutron-rich species.  The broad fan of the
Blandford-Payne MHD wind that is conducive to a robust explosion thus
may also be counterproductive from a nucleosynthesis point of view.
There are several factors that may mitigate this production of
neutron-rich ejecta.

Hoop stresses and the large ambient pressure may restrict the
opening angle of the initial jet. This would decrease the
mass of neutron-rich matter directly ejected in the jet.  We
note that the jet should broaden as it passes though the
standing shock at about 200 km due to the sudden decrease
of the external pressure.  The external pressure drops by about an
order of magnitude at that point whereas the pressure of the plasma
confined within the jet remains the same.  The opening angle of
the jet might then expand by about the ratio of the sound speeds,
or a factor of about 3.  This may help to broaden the fan of
the wind, but at larger radii where matter is somewhat less
neutronized.  On the other hand, a broad jet fan may encompass all
the very neutron-rich matter originally expelled in an LW bubble
leading to expulsion of all that neutron-rich matter even if the LW
phase did not result in an explosion.

Another factor is that the jet is likely to go up the rotation
axis where the density will be relatively low due to the
effects of rotation.  After bounce, rotating models give densities at
100 km about a factor of 5 - 10 lower on the rotation axis
compared to the equator (e.g. Symbalisty 1984;
M\"onchmeyer, Sch\"afer, M\"uller \& Kates 1991).
That means for a given opening angle of the slow toroidal jet,
there will be relatively less matter to be intercepted by
about the same factor.  The attendant lower pressure will also
tend to promote a broader fan, so this must be checked
quantitatively.  The rotation will also hinder convection
in the equatorial direction and this may promote neutrino
emission up the rotation axis (Fryer \& Heger 2000).
This tendency may also help to decrease the density along the
rotation axis, the direction of propagation of the jet.

We also note that the complementary effect is a higher density
along the equator.  For extreme values of rotation, this matter
will form an accretion torus with an accretion time long compared
to the dynamical timescale.  The large rotation and distorted geometry of this 
protopulsar phase will evolve through shedding of angular momentum 
to the jet and environment to the more familiar, more nearly 
spherical, pulsar phase.

Further out in the star, the rotation will cause less of an effect on
the isodensity contours, but there will still be some distortion.
In spherical core collapse supernovae models, iron is produced and
ejected by the  action of the shock on the silicon layers and the
associated production of $^{56}$Ni that subsequently decays to form
$^{56}$Fe. After several seconds, this silicon will be part of the
infalling matter.  The jet will create bow shocks that converge on
the equator and drive overlying matter out preferentially in the
equatorial plane  (Khokhlov et al. 1999).
The bow shocks will thus tend to drive the silicon layer and the newly
formed nickel outward along the equator.  This effect would be enhanced
by any rotational distortion.  Some of the pre-existing silicon may be
directly
ejected in this way and some silicon will be produced and ejected by
shocking the oxygen layer.  Silicon produced in the explosion may thus
also be  blown off in an equatorially preferential manner, but at
somewhat  lower velocities.  These dynamics may
partially account for the observation of high
velocity iron lying outside lower velocity silicon within the body of
Cas A (Hughes et al. 2000; Hwang et al. 2000) and for high velocity
knots of silicon found beyond the reverse shock (Fesen \& Gunderson
1996).  The jets themselves should be formed of neutron-rich
iron-peak material, but how much of this matter escapes the star
will depend on the envelope structure and shock dynamics.

\subsection{The Magnetic Switch, Failed Supernovae, and the Gamma-Ray Burst
Jet}

An important aspect of the dynamics of the jet propagation through
a star is that the faster the jet, the less likely it is to
trigger an explosion.  According to the calculations of Khokhlov \&
H\"oflich (2001) and H\"oflich, Khokhlov \& Wang (2001) 
(see also MacFadyen, Woosley \& Heger 2001), faster jets
have narrower opening angles, allowing less time for transverse
shocks to affect the star before the jet emerges from the boundary of the
core.  The jet energy will tend to stay focused and directed along the
jet trajectory rather than powering the bow shock and its associated
transverse sheath of shocks that are critical for spreading the jet energy
throughout the mantle and exploding the star. (The difference is akin to
the result of aiming a baseball and a needle with the same energy
at a loaf of bread.)  This leads to a suggestion
for a different ultimate evolution for those conditions that
lead to especially fast jets, namely a failed supernova, collapse
to form a black hole, and the attendant possibility of creating a \grb.
These ideas are summarized schematically in Figure 2.

We do not fully understand the conditions of either magnetic field strength
or rotation of the progenitor core.  It is plausible that these properties
have a finite distribution such that a small fraction of progenitors
have particularly large values of magnetic field and rotation.  For a
sub-set 
of all the core collapse events, e.g. in a given mass range, some could have
properties that satisfy the criterion that the MHD luminosity
exceeds the critical luminosity of the magnetic switch,
$L_{MHD} > L_{crit}$ (\S 4.2.2).  In this case, we expect the generation of
an
especially fast, narrow, jet with $v_{jet} \sim v_{A}$, which may
be well above $v_{esc}$.  Such a jet may be
less effective in creating a jet-induced supernova explosion.  If this is
the case, the mantle will continue to collapse
and a black hole would form.  The extreme version of this suggestion
is that the narrow jet fails to form any supernova.  It is plausible,
however, 
that some matter is expelled to trigger a supernovae, but an insufficient
fraction of the matter is ejected to prevent the ultimate fallback
of enough mass to form a black hole.

The delayed formation of a black hole would give a new ambience for
the generation of a third, very possibly highly relativistic, jet,
as widely discussed in
the literature (MacFadyen \& Woosley 1999; Aloy, et al. 2000;
MacFadyen, Woosley \& Heger 2001; Zhang \& Woosley 2001).  Whether
this relativistic jet can induce a supernova explosion when the
earlier, slower jet failed to do so is an open question, but as
we remarked above, the first jet could have generated a supernova
through {\em partial} ejection of the mantle
and still have allowed the formation of a black hole. In this case,
the highly-relativistic jet would form within an already exploding
configuration, an  assumption made in some of the computations (e.g.
Aloy et al.).  

What does seem clear is that if a new, relativistic jet can propagate
out of the infalling matter, it will collide with the iron-rich matter
expelled in the earlier jet.  Depending on the timescale for the
fallback to trigger the collapse to form the black hole (MacFadyen,
Woosley \& Heger 2000), the initial jet could have propagated well away
from the star.  For a typical velocity of the fast, but
mildly-relativistic protopulsar jet of $v \sim 10^{10}$ \cms, the
matter will have reached a distance of order $D \sim 10^{12}~t_{BH,2}$
cm, where $t_{BH,2}$ is the delay time between the initial collapse and
the formation of the black hole and relativistic jet in units of 100
s.  The material of the protopulsar jet may thus act as the ``target''
for the collision of the second jet and this collision could create the
fireball and the internal shocks that represent the onset of the \grb.
The interaction between the \grb\ jet and the target would occur at a
distance $D~c/(c-v)$, which is of order $D$, but
could be significantly greater as $v \rightarrow c$ and the \grb\ jet
takes longer to catch up to the fast protopulsar jet.  

In the observer frame, the duration of the \grb\ would be associated
with the time for the relativistic \grb\ jet to propagate along the
length of the mildly-relativistic jet, and this would be determined by
the longer of two time scales:  the duration of the non-relativistic,
protopulsar jet (since this determines the length of the iron-rich lobe
that is targeted) and the duration of the \grb\ jet itself.  The
former time scale will be approximately the spindown time of the
jet-emitting protopulsar:
\begin{equation}
\label{spindown}
\tau_{spindown} \simeq E_{rot,NS}/L_{MHD} \simeq 8~s\; B_{15}^{-2}
~ P_{ms}^{-1}~M_{NS}~R_{NS,6}^{-1}.
\end{equation}
In this scenario where the tripping of the magnetic switch triggers an
especially fast, inefficient jet, the field must exceed the critical
value (eqn. [\ref{crit-locus}]), so the spindown time will be
\begin{equation}
\label{spindown-crit}
\tau_{spindown} \lta 0.02~{\rm s}~P_{ms}^{-2}~\rho_{11}^{-1}~f_{\Delta\Omega}^{-1}
~M_{NS}^{-1/2}~R_{NS,6}^{3/2}.
\end{equation}
This spindown time can be as long as a typical \grb, $\sim 30$ s, 
for rotation at less than breakup, tight collimation, and appropriate
choice of density, so this could be the controlling factor in the
duration of a \grb\ if this scenario of colliding jets proves to
have merit. 

Using a strong magnetic field as the prime carrier of momentum and 
energy in the second, relativistic jet, with the jet material just 
along for the ride, might explain how high Lorentz factors can be
achieved and sustained in the black-hole-collapse jet.
For example, for a ``tenuous'' corona around the accreting
black hole of $10^6$ g cm$^{-3}$ in a magnetosphere of
$10^{16}$ G, equations (\ref{alfvenspeed})
and (\ref{equipspeed}) give relativistic Alfv\'en/jet speeds of
order
\begin{equation}
\label{alfvengamma}
\Gamma_{A} \approx \left(\frac{B^2}{4\pi\rho~c^2}\right)^{1/2} \simeq
100~B_{16}~\rho_{6}^{-1/2}.
\end{equation}
The corresponding power from the Blandford-Znajek mechanism from
a rapidly-rotating hole (Blandford \& Znajek 1977; Meier 2001) is:
\begin{equation}
\label{L-MHDBH}
L_{GRB} \simeq 10^{52} \, {\rm erg \, s}^{-1} \epsilon B_{16}^2 \,
M_{BH,0.5}^2,
\end{equation}
where $\epsilon$ is an efficiency factor and 
$M_{BH,0.5}$ is the black hole mass in units of 3 \m. 
If this luminosity is
pumped into the relativistic jet for a time $\tau_{rel}$ seconds, the total 
energy emitted in the jet is:
\begin{equation}
\label{E-GRB}
E_{GRB} \simeq L_{GRB} \, \tau_{rel} \simeq
10^{52} \, {\rm erg} \, \epsilon \, B_{16}^2 \, M_{BH,0.5}^2 \, \tau_{rel}.
\end{equation}
Recent analyses suggest that the typical energy emitted in
\grbs\ is around $3 \times 10^{50}$ ergs (Kumar \& Panaitescu 2001; 
Frail et al. 2001).  If this low energy is typical, then
very little of the expected black hole binding energy or rotation
energy can be tapped to make \grbs.  For the picture we sketch here and, 
indeed, for any picture based on the production of jets from black holes we must
have $ \epsilon B_{16}^2 \tau_{rel} \sim 0.03$.  In our particular
magnetic switch scenario, we must satisfy equation (\ref{crit-locus})
for the critical switch field and equation (\ref{alfvengamma}) to
produce a suitably relativistic jet.  This suggests that either
the efficiency for the Blandford-Znajek process is relatively low
(perhaps because the field is swallowed by the black hole rather
than producing electromagnetic flux) or the duration of the
process is short.  We note that if the latter is the case, then
the duration of the first jet (eqn. [\ref{spindown-crit}]) will
control the duration of the \grb.

Using the slower jet as the target and tying the delayed
interaction between the two jets to the mantle collapse time
would explain how \grs\ are produced so efficiently and routinely
far from the central black hole environment.  In addition, the
iron-rich nature of the first jet may give appropriate conditions
in which to address the iron lines observed in some \grb\ events
(e. g. Piro, et al. 2000).  The strength of the iron emission
is difficult to assess.  It depends on numerous very uncertain factors 
such as the mass of the iron ejected, the density in the precursor jet, 
the cross
sections of the first and second jets, and the ambient ionizing
flux.  Estimates in the literature of the amount of iron required to 
produce a given line flux by recombination vary by orders of magnitude 
(e.g. Piro et al. 2000; Rees \& M\'esz\'aros 2000; 
M\'esz\'aros \& Rees 2001).   

This scenario of jet collisions does not help
to resolve the issue of the low afterglow densities
associated with some \grbs\ (Kumar \& Panaitescu 2001; Scalo
\& Wheeler 2002b) that are difficult to achieve in any
model based on a massive star that expels a substantial wind.
It also offers no explanation as to why the
blast energy associated with the \grb\ is confined to a narrow
and rather small value, around $3 \times 10^{50}$ ergs.
The question of why such a small fraction of the
available energy would be tapped in the jet is an important
problem for the future.

\section{Summary and Discussion}

We have presented the case that strong toroidal magnetic fields
will be generated in stellar collapse and that these magnetic
fields can generate magneto-centrifugal jets in analogy to
those found in simulations of black hole accretion.
The case for magneto-centrifugal jets suggests
they may be frequent and robust and hence provide a good
basic understanding of why all core collapse supernovae are
found to be substantially asymmetric and predominantly bi-polar.

There are concerns that the jets we describe arise deep
in the gravitational potential where the infalling matter
is very neutron rich.  We argue that this constraint rules
out the prompt LeBlanc-Wilson jets as the common origin of
supernovae.  This suggests that conditions of initial rotation
and magnetic field extreme enough to produce strong jets in
this phase do not arise frequently in nature.  We point out
that successful jets from the later, protopulsar phase must
walk the line between being broad enough
to provide ample energy to the overlying matter without being
so broad as to eject excessive neutron-rich matter.  Partial
evacuation of the matter on the axis by rotation may alleviate
this problem, but it remains a concern.  The jets will produce
bow shocks that tend to expel matter, including iron and silicon,
in equatorial tori.  This may help to account for
observations of the element distribution in Cas A.

There also is some concern that gravitational radiation may remove so
much angular momentum, and so quickly, that there would not be enough
rotational energy to power a substantial MHD jet.  We estimate
that the effects of gravitational radiation (GR) will be significant,
but not severe.  The most important GR mode is probably the $r$-mode.
Detailed computations by Lindblom, Tohline, and Vallisneri (2001) on
the evolution and effects of this mode estimate that only about half of
the rotational kinetic energy will be lost before the gravitational 
radiation process
switches off.  The time scale for this to occur may be fairly short (a
few tens of initial rotational periods, i.e. as short as, say, 30 ms),
but the protoneutron star still will be left with a significant
fraction of its initial rotational energy, which can be tapped for the
production of MHD jets.

The topics of core-collapse supernovae and \grbs\ overlapped, at least
in principle,  with the discovery of SN 1998bw.
Although the association is still controversial,
this odd supernova is likely to have been connected to
GRB~980425 (Galama et al. 1998).  Supernova-like
excesses of light have been detected in the afterglow of
two \grbs, GRB~970228 (Reichart 1999, Galama et al. 2000) and GRB~980326
(Bloom et al. 1999) about two weeks after the \grbs.
The excess light in the afterglow of both GRB~970228 and
GRB~980326 has been modeled by the addition of light
from an event like SN~1998bw.  While there may be other
explanations for these excesses, the connection with
supernovae must be pursued.   We suggest here a new way to
make this connection.  The conditions
that apply in stellar collapse suggest that a magnetic ``switch''
mechanism found to apply in the black hole simulations may
also apply in the collapse case with subsequent affect on the
speed of the jet. If the switch turns on at low density and
large magnetic field, an especially fast jet, with propagation
speed of order the Alfv\'en speed, could be produced. This
jet would tend to propagate rapidly through the star and
deposit relatively little momentum.  The result might
be a supernova explosion, but with enough infall to produce
a delayed black hole.  A reprise of the physics outlined
here, and that explicitly invoked to produce relativistic MHD jets
from stellar mass and AGN black holes, could then produce
a relativistic jet that catches up to the first jet after
it has emerged from the star.  The interaction of these
two jets could plausibly be the origin of the internal
shocks thought to produce \grbs\ and could explain the presence of
iron lines in the afterglow.
  
In this paper we have discussed
toroidal fields of order $10^{14}$ to
$10^{16}$ G arising routinely in the formation of a neutron star
in a supernova.  There is clearly an issue of the
final effective dipole strength of the neutron stars
left behind in the explosion.  We must have a final
dipole field distribution of the neutron stars
consistent with those of observed pulsars,
$\sim$  $10^{12}$ to $10^{13}$ G.
It seems plausible that the large fields generated
in the collapse are dissipated in the helical jets.
The very strong fields that may be generated in the depths
of the neutron star will take longer to float out, but
the time still could be short compared to the lifetime of the
supernovae, never mind the pulsar.  Stronger fields in
the interior could also leave smaller surface dipole fields.
Much of the field may be amplified and dissipated at the shearing boundary
between the infalling matter and the neutron star.
This layer and its field will almost surely substantially
dissipate after the explosion.
These issues should be given more careful consideration.

All of the issues raised here should be examined in the context of
magnetars.  However the large effective dipole fields arise
in magnetars, there is a strong presumption that they are
created in the formation of the neutron star.  The formation
of a magnetar may require special circumstances, or they
may arise as the tail of the normal pulsar formation process.
One possibility is that they represent that fraction of core-collapse
events that, perhaps by accident of progenitor conditions, do
attain an especially rapid differential rotation and amplify the
field not just by field line wrapping, but by an $\alpha - \Omega$
dynamo (Duncan \& Thompson 1992).  This could lead to exponential
field growth, and total field strength of order $10^{17}$ G with dipole
component of order $10^{15}$ G. This is approximately the regime
of fast toroidal jets that we invoke to produce less ejecta and
more infall to produce a black hole and perhaps a \grb.  To
reconcile these pictures, it may be that magnetars arise with
somewhat higher than normal fields and rotation, but still in
the slow toroidal jet phase with conditions where the magnetic
switch is still ``off." Even more extreme conditions could
lead to the flipping of the switch, producing an especially fast
jet that would be less effective in driving the explosion
and thus leaving a black hole. Clearly there is a great need
to study the many aspects of this problem in numerical and
quantitative detail.

Some models of SN~1998bw invoked especially large kinetic energies,
in excess of $10^{52}$ ergs in spherically-symmetric models, to
account for the bright light curve and high velocities (Iwamoto, et al.
1998; Woosley, Eastman, \& Schmidt 1998), while others took note of
the measured polarization to suggest that strongly asymmetric
models could account for the observations with more ``normal''
energies (H\"oflich, Wheeler, \& Wang 1999).  Large kinetic
energy has also been attributed to several other supernovae
(again based on spherically symmetric models), especially
SN 1997cy (Germany et al. 2000) and SN 1997ef (Branch 2001;
Iwamoto et al. 2000).  Events like SN~1997cy, SN~1997ef and SN~1998bw
will help to sort out the physics of explosive events, whether such
events are more closely related to ``ordinary'' supernovae or
``hypernovae'', whether either of these classes leaves behind
neutron stars as ``ordinary'' pulsars or highly magnetized
``magnetars'' or whether the
remnant is a black hole and whether any of these events are associated
with classic cosmic \grbs\ as suggested by the supernova-like
brightening of the afterglow of GRB~970228 and GRB~980326.
In terms of our magnetic switch mechanism, it may be that
``hypernovae" are still in the switch ``off" phase so they
have strong jets, but not so fast and narrow that the explosion
is mitigated.  In this sense, there may be a connection between
the ``hypernovae" and magnetars that, as argued above, may
arise in the same regime of especially energetic protopulsar jets,
but still with the switch ``off."

If it were not for the constraints of the supernova polarimetry,
we might continue to study essentially spherically symmetric
collapse models, as indeed, many people will and must.
It may be that we are pessimistic and that neutrino asymmetries
can account for the observed polarimetry.  The fact that jets alone can
account for supernova explosions and the observed asymmetries
has been established in principle.  We think
nature is saying that jets must occur in routine supernovae.  We also
believe that the
process of neutron star formation must inevitably involve
the physics outlined here whether the resulting jets are weak
or strong.  
Clearly, we need a more rigorous
description of the process of field generation,
buoyancy, emergence, and subsequent evolution.

\section{Acknowledgments}
We are grateful to Alexei Khokhlov, Peter H\"oflich, Elaine Oran and
Almadena Chtchelkanova for helping us to understand
how jets work in stars, to Lifan Wang for showing that
core collapse supernovae are asymmetric, to Rob Duncan for discussions of
magnetars, and to Ellen Zweibel for discussion of MHD and plasma physics.
We would especially like to thank the Aspen Center for
Physics for the hospitality during the gestation of this collaboration.
This research was supported
in part by NSF Grant 95-28110, NASA Grant NAG 5-2888, a grant
from the Texas Advanced Research Program.
Some of this research was carried out at the Jet Propulsion Laboratory, 
California Institute of Technology, under contract to the National 
Aeronautics and Space Administration.

\vfill\eject

\noindent
{\bf References}

\noindent
Akiyama, S., Wheeler, J. C. \& Meier, D. L. 2002, in preparation

\noindent
Akiyama, S., Wheeler, J. C. \& Lichtenstadt, I, 2002. in preparation

\noindent
Aloy, M. A., M\"uller, E., Ib\'e\~nez, J. M., Mart\'i, J. M.
\& MacFadyen, A. 2000, ApJ, 531, L119

\noindent
Bisnovatyi-Kogan, G. S. 1971, Soviet Astronomy AJ, 14, 652

\noindent
Bisnovatyi-Kogan, G. S. \& Ruzmaikin, A. A. 1976, Ap\&SS, 42, 401

\noindent
Blandford, R. D. \& Payne, D. G. 1982, MNRAS, 199, 833

\noindent
Blandford, R. D. \& Znajek, R. 1977, MNRAS, 179, 433


\noindent
Branch, D. 2000, in Supernovae and Gamma-Ray Bursts: the Greatest 
Explosions Since the Big Bang, eds. M. Livio, N. Panagia, \& K. Sahu
(Cambridge: Cambridge University Press). p. 96 

\noindent
Burrows, A. \& Hayes, J. 1996, Phys Rev Lett, 76, 352

\noindent
Burrows, A., Hayes, J. \&  Fryxell, B. A. 1995, ApJ, 450, 830

\noindent 
Cameron, A. G. W. 2001, ApJ, in press

\noindent
Chevalier, R. A. \& Soker, N. 1989,  ApJ, 341, 867 

\noindent
Duncan, R. C. \& Thompson, C. 1992, ApJ, 392, L9

\noindent
Fesen, R. A. \& Gunderson, K. S. 1996, ApJ, 470, 967

\noindent
Frail, D. A. et al. 2001, ApJ, submitted. 

\noindent
Fryer, C. L. \& Heger, A., 2000, ApJ, 541, 1033

\noindent
Galama, T. J. et al. 1998, Nature, 395, 670

\noindent
Galama, T. J. et al. 2000, ApJ, 536, 185 

\noindent
Germany, L. M., Reiss, D. J., Sadler, E. S., Schmidt, B. P. \& Stubbs, C. W.
2000, ApJ, 533, 320 


\noindent
Helfand, D. J. Gotthelf, E. V., \& Halpern, J. P. 2001, ApJ, 556, 380.

\noindent
H\"oflich, P., Khokhlov, A. \& Wang, L. 2001, in Proc. of the 20th
Texas Symposium on Relativistic Astrophysics, eds. J. C. Wheeler \&
H. Martel, (New York: AIP), 459 

\noindent
H\"oflich, P., Wheeler, J. C., \& Wang, L. 1999, ApJ, 521, 179


\noindent
Hughes, J. P., Rakowski, C. E., Burrows, D. N. \& Slane, P. O.
2000, ApJ, 528, L109

\noindent
Hwang, U. Holt, S. S. \& Petre, R. 2000, ApJ, 537, L119
 
\noindent
Ibrahim, A. I., Strohmayer, T. E., Woods, P. M., Kouveliotou, C.,
Thompson, C., Duncan, R. C., Dieters, S.,  van Paradijs, J. \& Finger, M.
2001, ApJ, 558, 237

\noindent
Iwamoto K. et al., 1998 Nature, 395, 672

\noindent
Iwamoto K., Nakamura, T., Nomoto, K., Mazzali, P. A., Danziger, I. J.,
Garnavich, P., Kirshner, R. P., Jha, S., Balam, D.
\& Thorstensen, J. 2000, ApJ, 534, 660

\noindent
Janka, H.-T. \& M\"uller, E. 1996, A\&A, 306, 167

\noindent
Khokhlov,  A. M., 1998, J. Comput. Phys., 143, 519

\noindent
Khokhlov, A. \&  H\"oflich, P. 2001, in Explosive Phenomena
in Astrophysical Compact Objects, eds. H.-Y, Chang, C.-H. Lee
\& M. Rho, AIP Conf. Proc. No. 556, (New York: AIP), p. 301

\noindent
Khokhlov,  A. M., H\"oflich P. A., Oran E. S., Wheeler J. C.
Wang, L, \& Chtchelkanova, A. Yu. 1999, ApJ, 524, L107

\noindent
Koide, S., Shibata, K. \& Kudoh, T. 1999, ApJ, 522, 727

\noindent
Koide, S., Meier, D. L., Shibata, K. \& Kudoh, T. 2000, ApJ, 536, 668

\noindent
Konigl, A. \& Pudritz, R. E. 2000, in Protostars and Planets IV,
eds. V. Mannings, A. Boss, \& S. Russell
(Arizona: University of Arizona Press), p. 759

\noindent
Kouveliotou, C. et al.  1999, ApJ, 510, L115

\noindent
Krasnopolsky, R., Li, Z., \& Blandford, R.\ 1999, ApJ, 526, 631

\noindent 
Kudoh, T., Matsumoto, R., \& Shibata, K.\ 1998, ApJ, 508, 186

\noindent
Kundt, W. 1976, Nature, 261, 673

\noindent
Lai, D., Chernoff, D. F. \& Cordes, J. M. 2000, ApJ, 549, 1111

\noindent
Lery, T. \& Frank, A.  2000, ApJ, 533, L897

\noindent
LeBlanc, J. M. \& Wilson, J. R. 1970, ApJ, 161, 541

\noindent
Leonard, D.~C., Filippenko, A.~V., Barth, A.~J., \& Matheson, T.\ 2000, ApJ,
536, 239 

\noindent
Leonard, D.~C., Filippenko, A.~V., Ardila, D.~R., \& Brotherton, M.~S.\
2001, ApJ, 553, 861

\noindent
Lindblom, L., Tohline, J.E., \& Vallisneri, M. 2001, Phys. Rev.
Letters, 86, 1152.



\noindent
Lovelace, R. V. E., Romanova, M. M. \& Bisnovatyi-Kogan, G. S.
1999, ApJ, 514, 368 

\noindent
MacFadyen, A. \& Woosley, S. E. 1999, ApJ,  524, 262

\noindent
MacFadyen, A., Woosley, S. E., \& Heger, A. 2001, ApJ,  550, 410

\noindent
Meier, D. 1999, ApJ, 522, 753

\noindent
Meier, D. 2001, ApJ, 548, L9.

\noindent
Meier, D., Edgington, S., Godon, P., Payne, D. G. \& Lind, K. R. 1997,
Nature, 388, 350   

\noindent
Meier, D., Epstein, R. I., Arnett, W. D. \& Schramm, D. N. 1976,
ApJ, 204, 869   

\noindent
M\'esz\'aros, P. \& Rees, M. J. 2001, ApJ, 556, L37

\noindent
Mezzacappa, A. et al. 1998, ApJ, 495, 911.

\noindent
M\"onchmeyer, R., Sch\"afer, G., M\"uller, E. \& Kates, R. E. 1991, A\&A,
246, 417

\noindent
Nakamura, T. 1998, Prog. Theor. Phys. 100, 921

\noindent
Ostriker, J. P. \& Gunn, J. E. 1971, ApJ, 164, L95

\noindent
Ouyed, R., Pudritz, R. E. \& Stone, J. M. 1997, Nature, 385, 409


\noindent
Pun, C. S. J., et al. 2001, ApJ, submitted

\noindent
Rees, M. J. \& M\'esz\'aros, P. 2000, ApJ, 545, L73


\noindent
Romanova, M. M., Ustyuogova, G. V., Koldoba, A. V., Chechetkin, V. M.
\& Lovelace, R. V. E. 1998, ApJ, 500, 703

\noindent
Ruderman, M. A., Tao, L. \& Klu\'zniak, W. 2000, ApJ, 542, 243

\noindent
Scalo, J. M. \& Wheeler, J. C. 2002a, ApJ, in press

\noindent
Scalo, J. M. \& Wheeler, J. C. 2002b, ApJ, in press


\noindent
Shimizu, T., Yamada, S. \& Sato, K. 1994, ApJ, 432, L119

 

\noindent
Spruit, H. \& Phinney, E. S. 1998, Nature. 393, 139

\noindent
Symbalisty, E. M. D. 1984, ApJ, 285, 729

\noindent
Symbalisty, E. M. D., Schramm, D. N. \& Wilson, J. R. 1985, ApJ, 291, L11

\noindent
Uchida, Y., Nakamura, M., Hirose, S., \& Uemura, S.\ 1999, \apss, 264, 195

\noindent
Ustyugova, G. V., Koldoba, A. V., Romanova, M. M., Chechetkin, V. M. \&
Lovelace, R. V. E.
1999, ApJ, 516, 221

\noindent
Wang, L., et al. 2001b, ApJ, submitted 

\noindent
Wang, L., Howell, D. A., H\"oflich, P. \& Wheeler, J. C. 2001a, ApJ, 550,
1030 

\noindent
Wang, L., Wheeler, J. C., Li, Z. W., \& Clocchiatti, A. 1996, ApJ, 467, 435

\noindent
Weisskopf, M. C., et al. 2000, ApJ, 536, L81.

\noindent
Wheeler, J. C., Yi, I., H\"oflich, P. \& Wang, L. 2000, ApJ, 537, 810

\noindent
Wilson, J. R. \& Mayle, R. 1993, Phys. Rep. 227, 97

\noindent
Woosley S., Eastman R., Schmidt M. 1998, ApJ, 516, 788

\noindent
Zhang, W.~\& Woosley, S.~E.\ 2001, BAAS 198, 3803 
 
\vfill\eject 

\figcaption{Schematic illustration of the protopulsar or slow toroidal jet
phase.  The left panel (a blow-up of the very inner portion of the right panel)
shows the weak LeBlanc/Wilson jet emerging from 
deep within the newly born neutron star.  Also shown is the torus of magnetic
field that will be created by wrapping of the initial field that is anchored
in the outer core.  The wrapping leads to field amplification, buoyancy, and
emergence of an accelerating helix of field (solid lines) that expands, 
carrying plasma in jet-like flow driven by field torsion and confined
by field tension.  The right panel shows the effect of the jet on the surrounding
star.  Bipolar jets propagate up the rotation axis.  Bow shocks lead to the
lateral propagation of shocks that collide on the equator and drive preferential
motion outward there. The net effect is to drive a strongly asymmetric explosion
into the outer portions of the star.}

\figcaption{Schematic illustration of the effects of a fast toroidal jet.  Left 
panel (a blow-up of the very inner portion of the central panel) shows a tightly
collimated, narrow, fast jet that would occur for conditions when the ``magnetic
switch" was ``on" (see text).  The fast jet can break out of the star, so most
of the energy is dissipated along the jet, rather than in lateral shocks (see
Figure 1).  The result can be the failure to explode the star or a weakened 
explosion that results in continued collapse to form a black hole.  The
``failed supernova jet" will proceed to propagate outward, and the
extreme conditions of rotation and gravity in the environment of the black hole
can create an especially fast, highly-relativistic MHD jet (right panel).  The relativistic
jet can collide with the previously expelled matter of the ``failed supernova jet"
which acts as a target to create internal shocks and to produce a \grb.}

\begin{figure}
\centering
\epsfig{figure=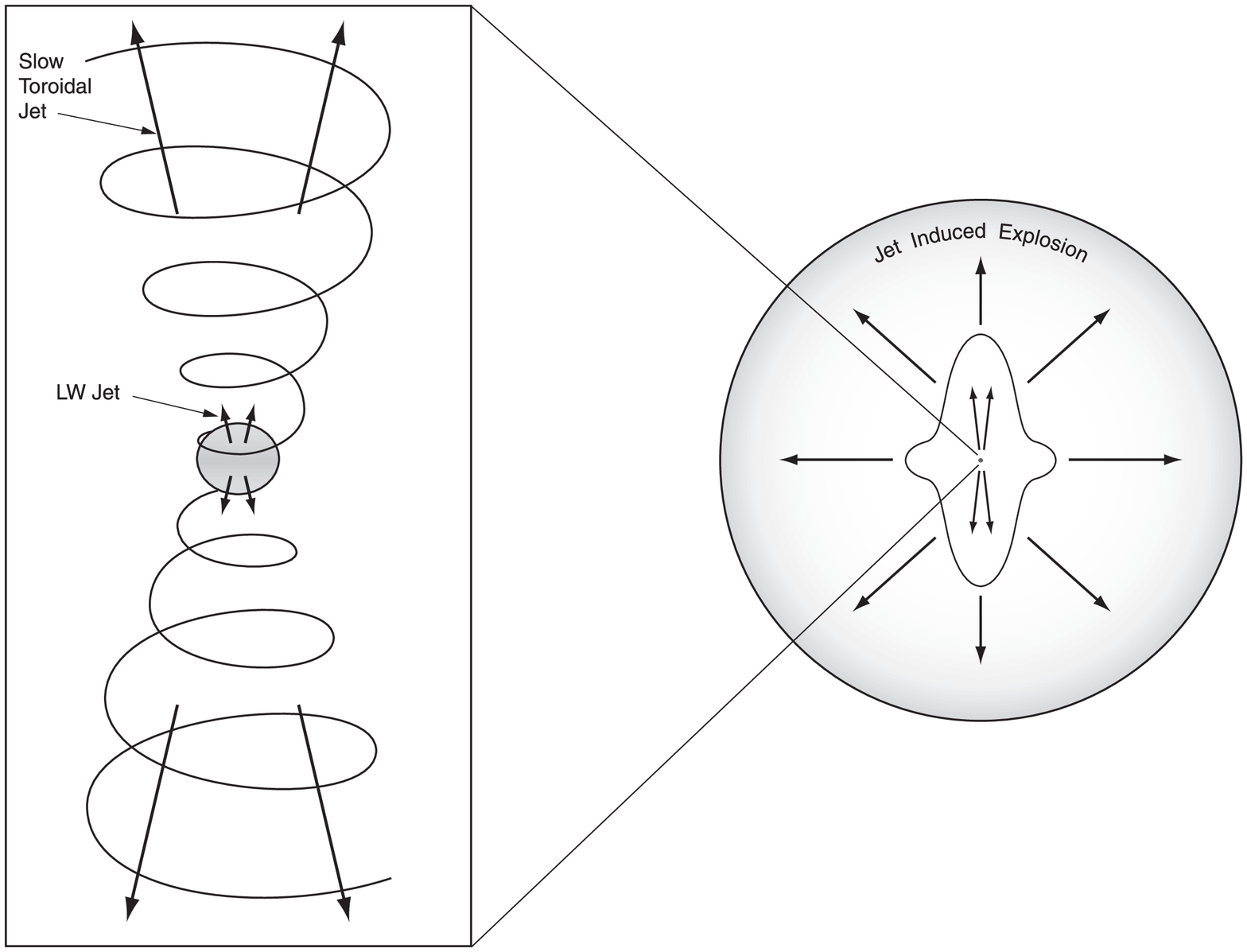,width=0.9\linewidth,angle=-90}
\end{figure}

\begin{figure}
\centering
\epsfig{figure=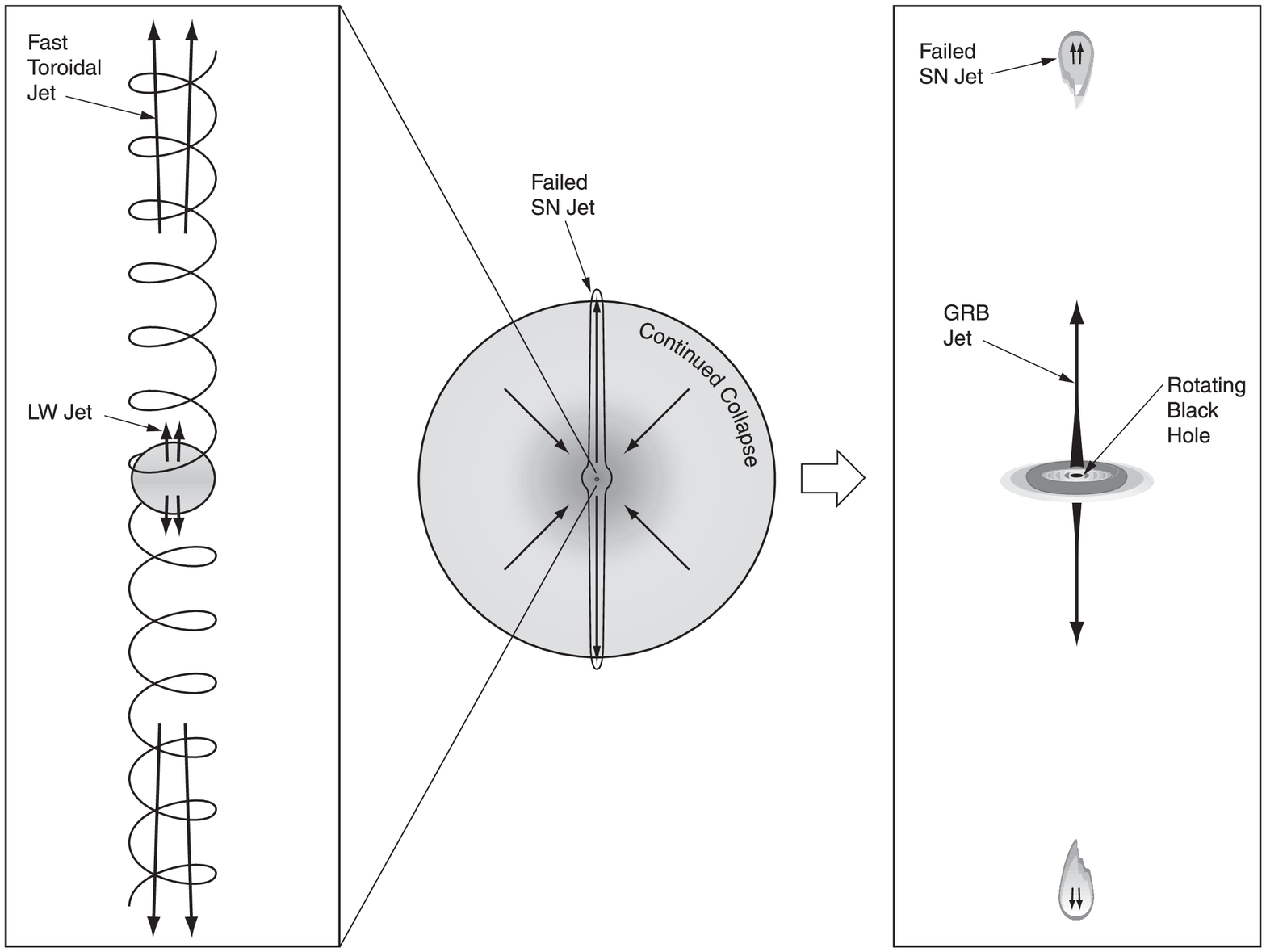,width=0.9\linewidth,angle=-90}
\end{figure}

\end{document}